# Time resolution of a Thick Gas Electron Multiplier (THGEM) - based detector

**Raz Alon**[*], **Marco Cortesi, Amos Breskin and Rachel Chechik**

*Department of Particle Physics,*
*Weizmann Institute of Science,*
*76100 Rehovot, Israel*

[*]*E-mail*: `raz.alon@weizmann.ac.il`

ABSTRACT: The time resolution of a double-stage Thick-GEM (THGEM) detector was measured with UV-photons and relativistic electrons. The photon detector, with semitransparent- or reflective-photocathode yielded time resolution of about 8-10ns RMS for single photoelectrons and 0.5-1ns RMS for few-hundred photoelectrons per photon-pulse. Time resolution of about 10ns RMS was recorded for relativistic electrons from a $^{106}$Ru source.

KEYWORDS: Thick Gas Electron Multiplier, Gas-avalanche, Time resolution

# Contents

## 1. Introduction

Contemporary and future experimental high energy physics poses ever-growing requirements for improved radiation detection technologies, with demands for better large-area, robust, position- and timing-sensitive detectors, at lower design and manufacturing cost. A promising newly developed detector is the THick Gaseous Electron Multiplier (THGEM) [1]. The THGEM is economically produced by drilling and etching a thin Cu-clad printed circuit board; electron multiplication occurs within sub-millimeter diameter holes, in a high dipole field established upon the application of a potential difference across the holes. The THGEM has been extensively investigated over the last 5 years [1-9]; most of its attributes are understood, as reviewed in [10]. In this paper we present new results on the time-resolution properties of a THGEM-based detector.

## 2. Methodology

A schematic view of a THGEM-based detector is shown in Figure 1. The measurements were carried out at atmospheric gas pressure, under a continuous flow of Ar/CH$_4$ (95:5). The electrodes were biased by independent high-voltage power supplies (CAEN N471A) through 20 MΩ resistors. Signals were recorded from the bottom electrode of THGEM2, keeping a reversed induction field of 200 V/cm (see Figure 1).

Measurements with UV photons were carried out with a spontaneously-discharging hydrogen lamp, providing multi-photon pulses. The detector elements, with an active area of 28x28 mm$^2$, were installed inside a stainless-steel vessel, equipped with a quartz window. The photons were converted on a CsI photocathode, which was either evaporated on a transparent quartz plate and installed at 3mm above the first THGEM (semi-transparent photocathode, not shown in Figure 1), or evaporated directly onto the top surface of the first THGEM (reflective photocathode, Figure 1 b) [2]. The THGEM geometry used was t=0.4mm, d=0.3mm, a=0.7mm, and the drift, transfer and induction gaps were 3mm, 4mm, and 3.6mm, respectively. The number of photons reaching the detector was controlled with absorbers inserted between the UV lamp and the window and could be reduced to a level of a single-photoelectron per pulse. The signals from the detector were amplified by a fast ESN VT110CH4 preamplifier, followed by an ORTEC 474 timing filter amplifier. They were further processed with ORTEC 934 constant-fraction discriminator and fed into a time-to-amplitude converter, along with the processed trigger signal from the discharge UV lamp. The time differences between the trigger signal and the detector signal were histogrammed using an Amptek 8000A multi-channel analyzer.

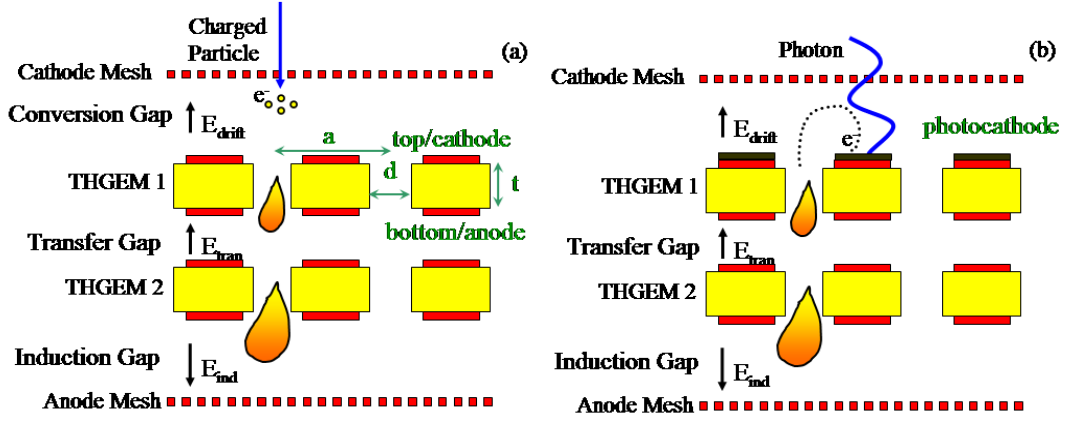

**Figure 1:** Schematic view of a double-THGEM detector; (a) conversion of incoming radiation in gas and (b) photon conversion on a reflective photocathode. The THGEM geometry is described by the thickness *t*, the hole-diameter *d* and the hole-pitch *a*.

The measurements with $^{106}$Ru electrons were carried out with a compact detector, of which the elements, with an active area of 28x28 mm$^2$, were assembled with epoxy glue. The trigger was provided by a 4.7mm thick plastic scintillator placed behind the detector and coupled to a Hamamatsu R6095 PMT. A 1.5mm G-10 absorber was placed between the detector and the PMT in order to bias the trigger towards the more energetic electrons in the spectrum. The detector was operated under continuous gas flow. The THGEM geometry was t=0.4mm, d=0.3mm, a=0.8mm. The conversion, transfer and induction gaps were all 3mm wide. Signals from the detector were amplified by a VV43 fast preamplifier (MPI – Heidelberg) followed by timing filter amplifier; the rest of the electronic chain was identical to that described above.

The electric fields distribution for the different detector configuration were calculated using MAXWELL 3D [10], and the electron drift-paths were simulated using GARFIELD [11]. These provided the distributions of electrons' arrival times, including the effect of diffusion, and provided an estimate of the expected time resolution for each configuration.

## 3. Results

We present results obtained with UV-photons, with both types of photocathodes, and with $^{106}$Ru electrons. Some simulation results are presented to be compared with the experimental ones. In what follows, σ is the standard deviation of the distribution of the time-difference between the detector and the trigger signals.

### 3.1 Timing with single- and multiple-photon pulses

The UV-lamp multi-photon pulses have a finite width, measured to be ~2ns RMS. When the intensity was attenuated to a single photoelectron/pulse, the trigger signal was randomly shifted in time with respect to the detected photon, by up to that amount; thus the detected time included a time-jitter contribution due to the lamp pulse-width. However, as this contribution was negligible compared to the measured resolution with single photoelectron, no correction was applied. For a large number of detected photons per pulse there was no such jitter effect because it is always the first detected photon (photoelectron) in a pulse that defines the detector's pulse rise and thus the timing; in this case, the lamp's pulse-width was not expected to influence the measured time resolution. The experimental results, measured with a detector gain of $\sim 10^4$ are shown in Figure 2.

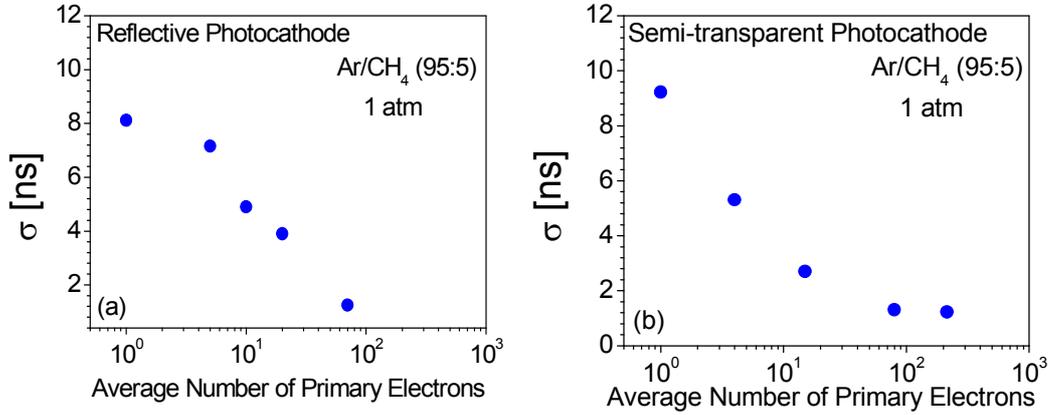

**Figure 2:** Measured time-resolution versus average number of primary photoelectrons in a double-THGEM photon detector (**Figure 1**) for (a) reflective CsI photocathode, and (b) semi-transparent photocathode.

At lower detector gain, of $\sim 10^3$, it was possible to further increase the number of photons, and consequently the number of primary photoelectrons per pulse; in these conditions the resolution further improved, e.g. to 0.54ns RMS with 2000 photoelectrons in a double-THGEM with a reflective photocathode.

Simulations conducted for a reflective-photocathode configuration with single photoelectrons yielded an estimated time-resolution of 4.9ns RMS [9]. This number is lower than the measured one, probably due to physical effects not included in the simulation such as signal rise-time fluctuations, cathode-surface non-homogeneity and electronic noise.

The drift field strength has an effect on the electron's drift-velocity and diffusion, and consequently should affect the time-resolution, with detectors having semitransparent photocathodes. Figure 3 shows the results measured with single-photoelectrons in a double-THGEM having a semi-transparent photocathode. Also shown in Figure 3 are calculated time-resolution results versus drift field.

A slight increase in time-resolution was observed for drift-fields increasing up to 1000 V/cm. This could be related to the decrease in electron drift-velocity with field, in this gas.

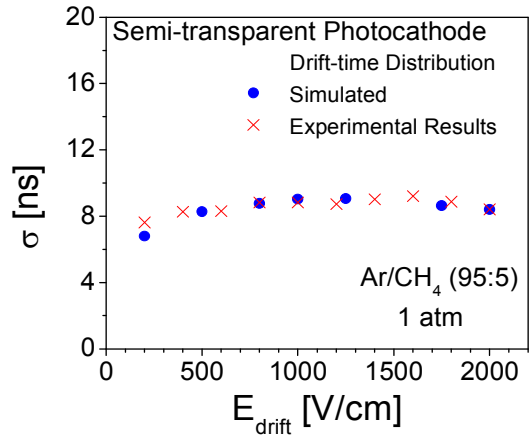

**Figure 3:** Experimental and calculated time-resolution values versus drift-field, in a double-THGEM with semi transparent photocathode operated with single photoelectrons.

The time resolution, measured with single-photoelectrons versus the transfer field between the two elements of the double-THGEM, is shown in Figure 4. The time resolution did not vary with the transfer field in the 0.5 to 3 kV/cm range.

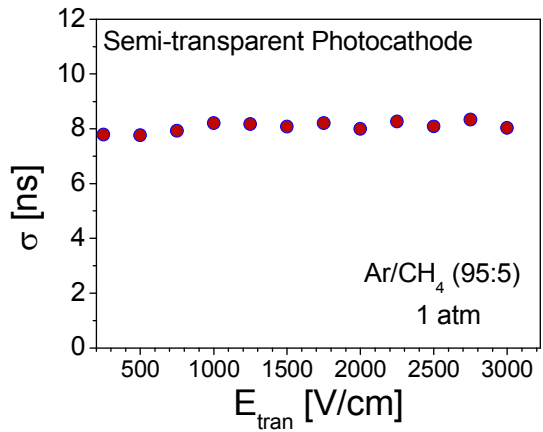

**Figure 4:** Experimental time-resolution versus transfer field, measured with single-photoelectrons in a double-THGEM with a semi-transparent photocathode.

### 3.2 Timing with relativistic electrons

Figure 5 shows the time-resolution measured with the $^{106}$Ru source; the measurements were carried out for various drift- and transfer field values. Similar to the measurements with single-photoelectrons, the time-resolution deteriorated when increasing the drift field, due to the corresponding decrease in electrons' drift velocity in this gas mixture. The effect of the transfer field was found to be minor over the range between 1 and 3 kV/cm.

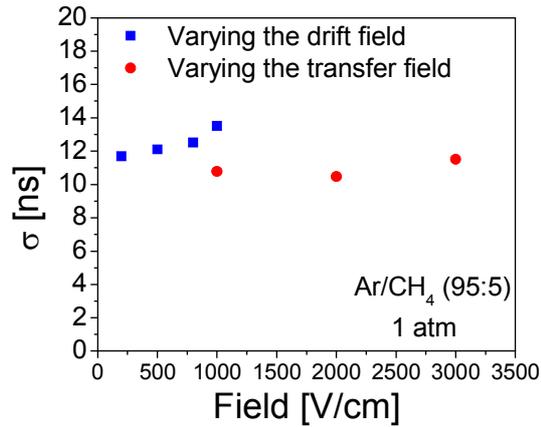

**Figure 5:** Experimental time-resolution measured with relativistic electrons versus drift- and transfer-field. The transfer-field was kept at 3 kV/cm while varying the drift field; the drift-field was kept at 0.5 kV/cm while varying the transfer-field.

In spite the fact that β-electrons from the $^{106}$Ru source deposit multiple ionization electrons along their path in the conversion gap, the measured time-resolution was of the same order as that measured with surface-deposited single-photoelectrons. We believe that this is a result of the exponential pulse-height distribution probability for each ionization electron. For the β-electron pulses (as for other relativistic charged particles) this resulted in strong pulse-to-pulse variations in rise-time, shape and amplitude (**Figure 6**), in addition to the statistical fluctuations of the deposition location of primary and secondary electrons within the conversion gap. The large fluctuations in pulse-shape made it very difficult to adjust the parameters of the constant-fraction discriminator, at the present low event rates, which strongly affected the time-resolution.

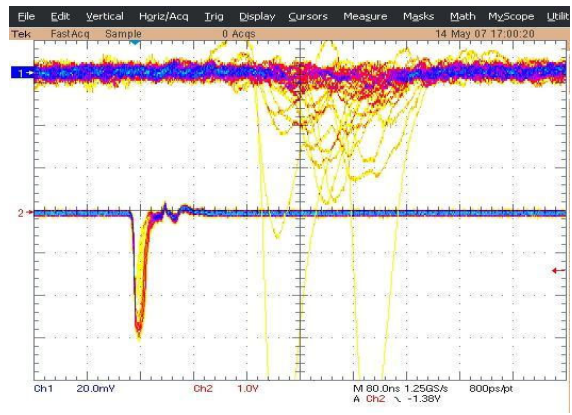

**Figure 6:** Fast pulses from the double THGEM detector (**Figure 1** a) recorded with β electrons (top), showing large pulse-to-pulse variations in shape and rise-time. The trigger PMT analog signal is shown at the bottom.

## 4. Summary

The time-resolution of a double-THGEM detector was studied with single and multiple UV-photons and with relativistic electrons from a $^{106}$Ru source. Resolutions of about 8ns RMS were reached with single-photoelectrons, with both semi-transparent and reflective CsI photocathodes. The resolution improved with increasing number of photoelectrons per pulse, reaching 0.54ns RMS for over a thousand photoelectrons. The improvement of the time-resolution with increasing number of photoelectrons per pulse originates from the increasing chance of first-arriving high-amplitude photoelectron pulses (better signal-to-noise ratio) to trigger the electronics above threshold. A slight deterioration of the time-resolution was observed with increasing drift-field, due to the behaviour of the electrons' drift velocity in the Ar/CH$_4$ (95:5) gas mixture; there was practically no variation with the transfer-field in the range of 1-3 kV/cm. Measurements with the $^{106}$Ru source proved to be difficult due to the strong pulse-to-pulse variations in pulse-height and shape of the signals induced by the β-electrons in the detector. The time resolution obtained with a double-THGEM detector was 10ns RMS. Better time resolution may be obtained with a mono-energetic beam from an accelerator or with a faster gas-mixture. Time resolution of about 8-9 ns RMS was measured with the $^{106}$Ru source and a triple-GEM detector operated in similar conditions (gas, fields, gain) [9]; this is to be compared with 5-7ns RMS time resolution reached with such detectors in particle-beams, though with different gas mixtures [13-14].
More details of the time resolution measurement can be found in [9].

In addition to the many attractive properties, such as high gain, good rate capability, sub-mm localization resolutions, low production cost per unit area and robustness, we showed that the THGEM offers relatively good time resolution for photons and relativistic charged particles.

## Acknowledgments

This work was partly supported by the Israel Science Foundation, grant No. 402/05 and by the MINERVA Foundation, grant No. 8566. A. Breskin is the W.P. Reuther Professor of Research in The Peaceful Use of Atomic Energy.